\documentclass[aps,prl,amsmath,amssymb,reprint,superscriptaddress,preprintnumbers,showpacs,intlimits]{revtex4-1}
\usepackage{bm,latexsym,mathrsfs,enumerate,color}
\usepackage[mathcal]{euscript}
\usepackage[breaklinks=true,unicode=true,urlcolor = blue,colorlinks = true,citecolor = blue,linkcolor = blue]{hyperref}
\usepackage{graphicx}
\usepackage{pgf}
\usepackage{tikz}
\usetikzlibrary{calc,intersections}
\graphicspath{{./figs/}}
\renewcommand{\vec}[1]{\bm{#1}}
\begin{document}

\title{Curvature effects in statics and dynamics of a thin magnetic shell}

\author{Yuri Gaididei}
 \affiliation{Bogolyubov Institute for Theoretical Physics, 03143 Kiev, Ukraine}

\author{Volodymyr P. Kravchuk}
 \email{vkravchuk@bitp.kiev.ua}
 \affiliation{Bogolyubov Institute for Theoretical Physics, 03143 Kiev, Ukraine}

\author{Denis D. Sheka}
\affiliation{Taras Shevchenko National University of Kiev, 01601 Kiev, Ukraine}

\date{\today}

%
%

\begin{abstract}
Equations of the magnetization dynamics are derived for an arbitrary curved 2D surface. General static solutions are obtained in the limit of a strong anisotropy of both signs (easy-surface and easy-normal cases). It is shown that the effect of the curvature can be treated as appearance of an effective magnetic field which is aligned along the surface normal for the case of easy-surface anisotropy and it is tangential to the surface for the case of easy-normal anisotropy. In general, the existence of such a field denies the solutions strictly tangential as well as strictly normal to the surface. As an example we consider static equilibrium solutions and linear dynamics for a cone surface magnetization.

\end{abstract}


\maketitle

Recent advances in microstructuring technology have made it possible to fabricate various low-dimensional systems with complicated geometry. Examples are cylindrical high-mobility two-dimensional (2D)  electron structures obtained by rolling-up mismatched semiconductor layers \cite{Schmidt01a}, flexible electronic devices \cite{Yuan06} and integrated circuits \cite{Kim08a}, spin-wave interference in rolled-up ferromagnetic microtubes \cite{Balhorn10}, magnetically capped rolled-up nanomembranes \cite{Streubel12b}  etc. After the seminal work of da Costa \cite{Costa81} where an effective  Schr{\"o}dinger equation for the tangential motion of a particle rigidly bounded to a surface was derived and the presence  of effective surface potentials  depending both on the Gaussian and mean curvatures was shown, much work has been done elucidating curvature effects in charge and energy transport and localization in systems with complicated geometry \cite{Burgess93,*Gaididei00a,*Entin02,*Gorria04,*Gaididei05a,*Ferrari08,*Jensen09,*Cuoghi09,*Korte09,*Bittner13}.
The behavior of  vector and tensor fields on curved surfaces has attracted attention of many researchers  (see e.~g. review papers \cite{Bowick09,Turner10}). However,  despite much work has been done it is not fully understood. One of the reasons for this is a complicate and intimate relation between two geometries: the geometry of the field (director in liquid crystalline phases, magnetization vector in ferromagnets, displacement vector in crystalline monolayer, etc) and the geometry of the underlying substrate. Until now researchers in this area were mostly concerned with the case when the vector field is strictly tangential to the curved surface, i.~e. 2D vector fields. This approach showed its validity and robustness in understanding crystalline arrangements of particles interacting on a curved surface \cite{Bowick09,Bowick00,Bowick01}, in studying   geometric interaction between defects and curvature in thin layers of superfluids, superconductors, and liquid crystals deposited on curved surfaces \cite{Vitelli04} and frustrated  nematic order in  spherical  geometries \cite{Lopez-Leon11}. However, the tangentiality  condition may be too restrictive for magnetic systems with their different types and strength of surface anisotropy (in/out of surface). Moreover, in the frame 2D vector field approach it is impossible to study  the dynamical properties of magnets on curved surfaces.

The goal of this Letter is to develop a full three dimensional (3D) approach to the static and dynamic properties of thin magnetic shells of arbitrary shape.


We base our study on the classical Landau-Lifshitz equation $\dot{\vec m}=\left[\vec m\times{\delta\mathcal{E}}/{\delta\vec m}\right]$, where $\vec m=\vec M/M_s$ is normalized magnetization unit vector with $M_s$ being the saturation magnetization, $\mathcal{E}=E/(4\pi M_s^2)$ is  normalized energy, the overdot indicates the derivative with respect to rescaled time in units of $(4\pi\gamma M_s)^{-1}$, $\gamma$ is gyromagnetic ratio. Since the dynamics of the vector $\vec m$ is precessional one, the energy functional $\mathcal{E}$ must be written for the case of general, \emph{not necessary tangential} magnetization distribution. Note, in case of static tangential distribution of the director in a curvilinear nematic shell the general expression for the surface energy was recently obtained in Refs.~\cite{Napoli12,Napoli12a}. The expressions for $\mathcal{E}$ for an arbitrary three dimensional magnetization distribution was already obtained only for cylindrical \cite{Landeros10,Gonzalez10} and spherical \cite{Kravchuk12a} geometries. Here we propose a general approach which can be used for an arbitrary curvilinear surface and an arbitrary magnetization vector field. However we neglect dipole-dipole interaction and take into account only exchange and anisotropy contributions. The last one can have a symmetry of the surface, e.g. it can be uniaxial with the axis oriented along the surface normal.

First of all we define a set of geometrical parameters of a curvilinear surface which will affect on the physical properties of the magnetic system.
Considering a 2D surface $\mathcal{S}$ embedded in 3D space $\mathbb{R}^3$, we use its parametric representation of general form $\vec{r}  = \vec{r}( \xi_1, \xi_2)$, where $\vec{r}  = x_i \hat{\vec{x}}_i$ is the 3D position vector defined in Cartesian basis $\hat{\vec x}_i\in\{\hat{\vec x},\,\hat{\vec y},\,\hat{\vec z}\}$, and $\xi_\alpha$ are local curvilinear coordinates on the surface.
Here and below Latin indices $i,j=1,2,3$ describe Cartesian coordinates and Cartesian components of vector fields, whereas Greek indices $\alpha$, $\beta=1,2$ numerate curvilinear coordinates and curvilinear components of vector fields. We also use here the Einstein summation convention.

Let us introduce the local normalized curvilinear basis ${\vec e}_\alpha={\vec g_\alpha}/{| \vec g_\alpha|}$, ${\vec n}=[{\vec e}_1\times {\vec e}_2]$, where $\vec{g}_\alpha = \partial_\alpha\vec{r}$ with $\partial_\alpha=\partial/\partial\xi_\alpha$. All the following analysis is made under an assumption that the basis  is orthogonal one or, equivalently, that the metric tensor $g_{\alpha\beta} = \vec{g}_\alpha \cdot \vec{g}_\beta$ is diagonal one. In other words, we choose the parametric definition of the given surface $\mathcal{S}$ in a form which provides orthogonality of the basis. For purpose of convenience of the further discussion we introduce vector $\vec\varpi$ of the spin connection $\varpi_\alpha=({\vec e}_1\cdot\partial_\alpha{\vec e}_2)$, the second fundamental form $b_{\alpha\beta} = {\vec{n}}\cdot \partial_\beta\vec{g}_\alpha$, and matrix $||h_{\alpha\beta}||=||b_{\alpha\beta}/\sqrt{g_{\alpha\alpha}g_{\beta\beta}}||$ which has the properties of the Hessian matrix:
the Gauss curvature $\mathcal{K}=\det(h_{\alpha\beta})$, the mean curvature $\mathcal{H}=\mathrm{Tr}(h_{\alpha\beta})/2$. It is instructive to emphasize that $g_{12}=g_{21}=0$ due to the orthogonality of the local basis and $b_{12}=b_{21}$ as well as $h_{12}=h_{21}$ by the definition.

Physically realizable magnetic nanomembranes are of finite thickness $L$.  We model such a nanomembrane as a thin shell with $L\ll\mathcal{R}$ with $\mathcal{R}$ being the minimal curvature radius of the surface $\mathcal{S}$. Then the space domain filled by the shell can be parameterized as $\mathfrak{r}(\xi_1,\xi_2,\eta)=\vec r(\xi_1,\xi_2)+\eta\vec{n}(\xi_1,\xi_2)$, where $\eta\in[-L/2,\,L/2]$. The main assumption is that the thickness $L$ is small enough to ensure the magnetization uniformity along direction of the normal, i.e. we assume that $\vec m=\vec m(\xi_1,\xi_2)$. This assumption is appropriate for the cases when the thickness is much smaller than the characteristic magnetic length. Similarly to \cite{Napoli12,Napoli12a}, we derive an effective 2D magnetic energy of the shell as a limiting case $L\to0$ of the 3D model and considering only linear with respect to the thickness $L$ contributions to the magnetic energy. Finally we consider the surface magnetic energy in form
\begin{equation}\label{eq:Energy-gen}
\mathcal{E}=L\int_\mathcal{S}\left[\ell^2\mathscr{E}_{ex}+\lambda (\vec m\cdot\vec n)^2\right]\mathrm{d}\mathcal{S},
\end{equation}
where the integration is over the surface $\mathcal{S}$ with the surface element $\mathrm{d}\mathcal{S}=\sqrt{g}\mathrm{d}\xi_1\mathrm{d}\xi_2$ where $g=\det(g_{\alpha\beta})$. The second term in the integrand is the density of the anisotropy energy, it is of easy-surface or easy-normal type for cases $\lambda>0$ and $\lambda<0$ respectively, here $\lambda$ is the normalized anisotropy coefficient. The exchange energy density is presented by the first term, where $\ell=\sqrt{A/(4\pi M_s^2)}$ is the exchange length and $A$ is the exchange constant.

In a Cartesian frame of reference an exchange energy density $\mathscr{E}_{ex}=(\vec\nabla m_i)(\vec\nabla m_i).$ The Cartesian components of the magnetization vector $m_i$ are expressed in terms of the curvilinear components $m_\alpha$ and $m_n$ as follows $m_i=m_\alpha({\vec e}_\alpha\cdot\hat{\vec x}_i)+m_n({\vec{n}}\cdot\hat{\vec x}_i)$. Then we substitute this expression into $\mathscr{E}_{ex}$ and apply the gradient operator in its curvilinear form $\vec\nabla\equiv(g_{\alpha\alpha})^{-1/2}{\vec e}_\alpha\partial_\alpha$. Everywhere in the text below the $\nabla$-operator is used in its curvilinear sense.
To incorporate the constrain $|\vec m|=1$, we also use the angular parametrization
\begin{equation}\label{eq:angular_repres}
\vec{m} = \sin\theta\cos\phi\, {\vec{e}}_1 + \sin\theta\sin\phi\,{\vec{e}}_2 + \cos\theta\, {\vec{n}},
\end{equation}
where $\theta=\theta(\xi_1,\xi_2)$ is the colatitude and $\phi=\phi(\xi_1,\xi_2)$ is the azimuthal angle in the local frame of reference. Finally, in terms of $\theta$ and $\phi$ the exchange energy density $\mathscr{E}_{ex}$ reads
\begin{equation} \label{eq:main-formula}
\!\!\mathscr{E}_{ex}\!\! = \left[\vec{\nabla}\theta -\vec{\Gamma}(\phi)\right]^2 + \left[\sin\theta\left(\vec{\nabla}\!\phi-\vec{\Omega}\right)\! - \!\cos \theta \frac{\partial \vec{\Gamma}(\phi)}{\partial\phi}\right]^2\!\!\!\!\!.
\end{equation}
Here the vector $\vec\Omega=\left(\varpi_1/\sqrt{g_{11}},\,\varpi_2/\sqrt{g_{22}}\right)$ is a modified spin connection and vector $\vec\Gamma$ is determined as follows
\begin{equation}\label{eq:Gamma-def}
\vec{\Gamma}(\phi)\!=\!||h_{\alpha\beta}||\vec\tau(\phi)=\mathcal{H}\,\vec\tau(\phi)+\sqrt{\mathcal{H}^2-\mathcal{K}}\,\vec\tau(\upsilon-\phi),
\end{equation}
where $\vec\tau(\phi)=\cos\phi\vec e_1+\sin\phi\vec e_2$  and the angle $\upsilon$ is given by $\upsilon=\arctan\Big(2\,b_{12}\sqrt{g}/(g_{22}b_{11}-g_{11}b_{22})\Big)$.

Using the energy expression \eqref{eq:main-formula} one can analyze general static solutions for the case of a strong anisotropy. Let us first consider the case of \underline{easy-surface anisotropy} ($\lambda>0$) and let the anisotropy be strong enough to provide a quasitangential magnetization distribution, in other words $\theta=\pi/2+\vartheta$ with $\vartheta\ll1$. Then the total energy \eqref{eq:Energy-gen} can be expressed as
\begin{equation} \label{eq:En-easy-surf-all}
\begin{split}
&\mathcal{E}\approx L\int\left(\ell^2\mathscr{E}^\mathrm{t}+2\ell^2\mathrm{F}^\mathrm{t}\vartheta +\lambda\vartheta^2\right) \mathrm{d}\mathcal{S},\\
&\mathscr{E}^{\mathrm{t}}=\vec{\Gamma}^2+(\vec\nabla\phi-\vec\Omega)^2\!\!,\quad
\mathrm{F}^\mathrm{t}=\nabla\cdot\vec\Gamma+(\nabla\phi-\vec\Omega)\frac{\partial\vec\Gamma}{\partial\phi}\!,\!
\end{split}
\end{equation}
where $\mathscr{E}^{\mathrm{t}}$ is the energy density of a strictly tangential distribution ($\theta\equiv\pi/2$ or equivalently $m_n\equiv0$), and $\mathrm{F}^\mathrm{t}$ can be treated as amplitude of a curvature induced effective magnetic field oriented along the normal vector $\vec n$.

Minimization of the energy functional \eqref{eq:En-easy-surf-all} results in
\begin{equation}\label{eq:theta-surf}
\vartheta=-\frac{\ell^2}{\lambda}\mathrm{F}^\mathrm{t}(\phi)+\mathcal{O}\left(\frac{1}{\lambda^2}\right),
\end{equation}
where the equilibrium function $\phi$ is obtained as a solution of the equation $\delta\mathscr{E}^\mathrm{t}/\delta\phi=0$. Accordingly to \eqref{eq:theta-surf} the strictly tangential solution is realized only for a specific case $\mathrm{F}^\mathrm{t}(\phi)\equiv0$.

The expression analogous to $\mathscr{E}^{\mathrm{t}}$ was recently obtained in Refs.~\cite{Napoli12,Napoli12a} for the case of curvilinear nematic shells with purely tangential distribution of the director.
However, as follows from \eqref{eq:theta-surf} the purely tangential solutions are not possible in general case.

For the opposite case of the strong \underline{easy-normal anisotropy} ($\lambda<0$) one has two possibilities, namely $\theta=\vartheta$ or $\theta=\pi-\vartheta$ with $\vartheta\ll1$. In the first case the total energy \eqref{eq:Energy-gen} can be written as
\begin{equation} \label{eq:En-easy-norm-all}
\mathcal{E}\approx L\int\left(2\ell^2\vartheta\,\mathrm{F}^\mathrm{n}+|\lambda|\vartheta^2/2\right)\mathrm{d}\mathcal{S} +\mathrm{const},
\end{equation}
where $\mathrm{F}^\mathrm{n} =(\nabla\cdot h)\cdot\vec\tau+\vec\Omega\left(h\, \frac{\partial\vec\tau}{\partial\phi}\right)$ can be treated as amplitude of a curvature induced effective magnetic field oriented along vector $\vec\tau$, here $(\nabla\cdot h)_\alpha=\frac{1}{\sqrt{g}}\partial_\beta\left(h_{\beta\alpha}\sqrt{g/g_{\beta\beta}}\right)$ is tensor generalization of the divergence. Minimization of the energy functional \eqref{eq:En-easy-norm-all} leads to the solution
\begin{equation} \label{eq:easy-normal}
\vartheta=-\frac{2\ell^2}{|\lambda|}\mathrm{F}^\mathrm{n}(\phi)+\mathcal{O}\left(\!\!\frac{1}{\lambda^2}\!\!\right)\!\!,\;
\tan\phi=\frac{(\nabla\cdot h)_2-(h\,\vec\Omega)_1}{(\nabla\cdot h)_1+(h\,\vec\Omega)_2}\!.\!\!
\end{equation}
There are two equilibrium values of the azimuthal angle: $\phi$ and $\phi+\pi$. One should choose that solution which provides $\vartheta>0$. 

Similarly to the previous case a solution strictly normal to the surface is realized only for the specific case $\mathrm{F}^\mathrm{n}\equiv0$. For spherical and cylindrical surfaces this condition is satisfied.

\begin{figure}
\includegraphics[width=\columnwidth]{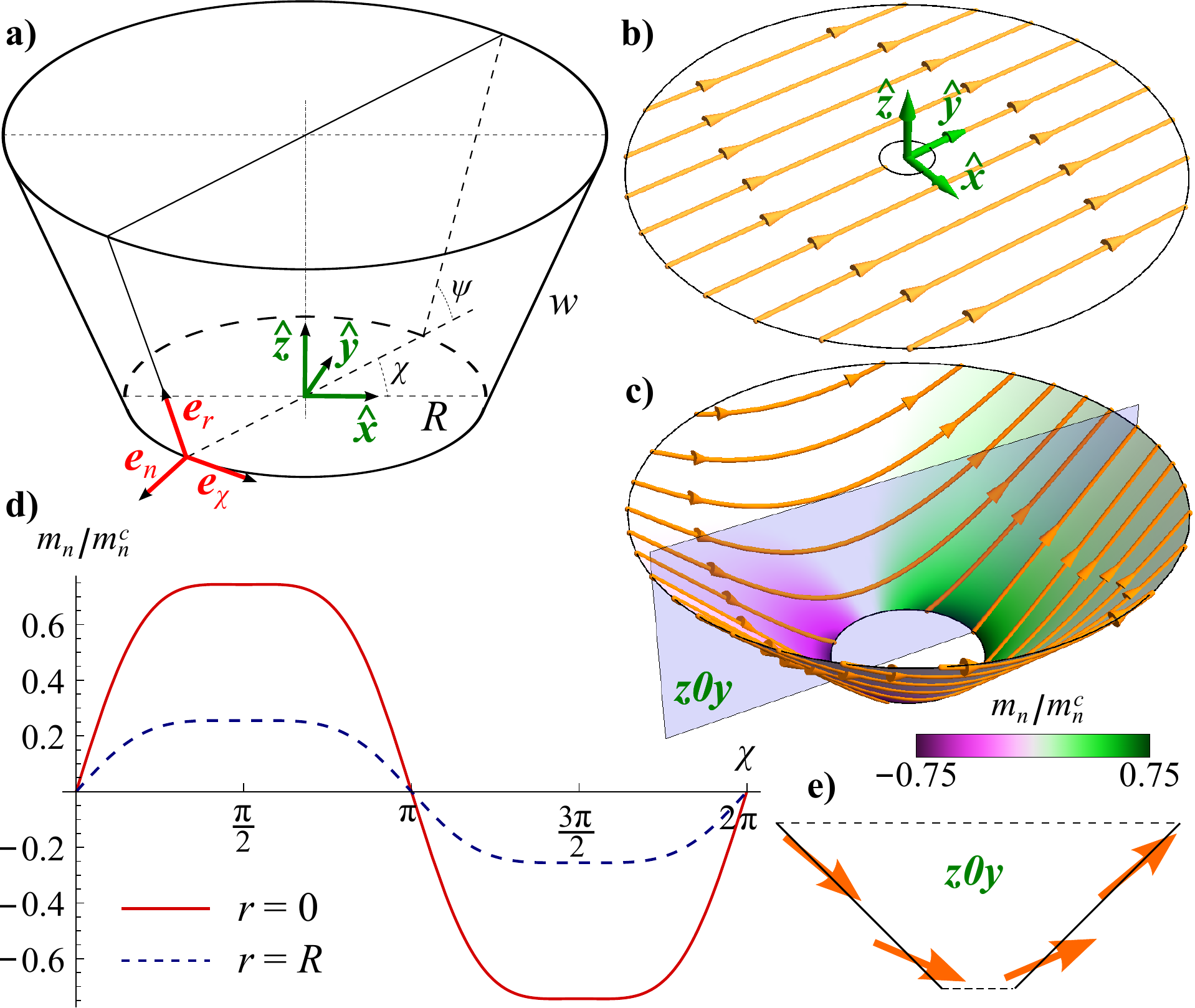}
\caption{(Color online) Onion state of a cone surface. Geometry of he problem and notations are shown on the inset a). Insets b) and c) demonstrate the onion solution \eqref{eq:onion} for cases $\psi=0$ and $\psi=\pi/4$ respectively. The streamlines demonstrate the in-surface magnetization distribution and the normal component $m_n$ is shown by the color scheme, the normalizing constant $m_n^c=\ell^2/(R^2\lambda)$. Variation of $m_n$ along the azimuth direction $\vec e_\chi$ for a cone with $\psi=\pi/4$ is shown on the plot d). Inset e) schematically demonstrates the magnetization distribution within the cut plane $z0y$.}\label{fig:onion}
\end{figure}

As the \textbf{first example} of application of our theory we find possible equilibrium states of cone shells with high anisotropies of different types.  We consider here side surface of a right circular truncated cone. Radius of the truncation face is $R$ and length of the cone generatrix is $w$. Varying the generatrix inclination angle $0\le\psi\le\pi/2$ one can continuously proceed from planar ring ($\psi=0$) to the cylinder surface ($\psi=\pi/2$), see Fig.~\ref{fig:onion}.

We chose the following parametrization of the cone surface
\begin{equation} \label{eq:cone-param}
x+i y =(R+r\cos\psi)\exp(i\chi),\qquad z=r\sin\psi,
\end{equation}
where the curvilinear coordinates $\chi\in[0,2\pi)$ and $r\in[0,w]$ play roles of $\xi_1$ and $\xi_2$ respectivelly. Definition \eqref{eq:cone-param} generates the following geometrical properties of the surface: the metric tensor $\begin{Vmatrix}g_{\alpha\beta}\end{Vmatrix}=\mathrm{diag}(g;1)$, the modified spin connection ${\vec\Omega}={\vec e}_\chi\cos\psi/\sqrt{g}$, the second fundamental form $\begin{Vmatrix}b_{\alpha\beta}\end{Vmatrix}=\mathrm{diag}(-\sin\psi\sqrt{g};0)$, and the matrix $\begin{Vmatrix}h_{\alpha\beta}\end{Vmatrix}=\mathrm{diag}(-\sin\psi/\sqrt{g};0)$, where $\sqrt g=R+r\cos\psi$. In accordance to the definition \eqref{eq:Gamma-def} one obtains $\vec\Gamma=-{\vec e}_\chi\sin\psi\cos\phi/\sqrt{g}$.

Let us start with the \underline{easy-surface case}. The solution will consist of two steps: (i) first, by minimizing the energy $\mathscr{E}^\mathrm{t}$ we obtain the main tangential distribution $\phi$, and (ii) using the obtained solution we calculate corrections for the out-of-surface component \eqref{eq:theta-surf}. For the cone surface \eqref{eq:cone-param} the energy $\mathscr{E}^{\mathrm{t}}$ can be written as follows
\begin{equation}\label{eq:Eex-phi}
\mathscr{E}^\mathrm{t} =\frac{1}{g}\left[\sin^2\psi\cos^2\phi+(\partial_\chi\phi-\cos\psi)^2\right]+(\partial_r\phi)^2.
\end{equation}
Accordingly to \eqref{eq:Eex-phi}, one should conclude that $\phi=\phi(\chi)$ for reasons of the energy minimization.
%
The variation of the total energy \eqref{eq:Energy-gen} with the density \eqref{eq:Eex-phi} results in the pendulum equation
\begin{equation} \label{eq:phi}
\phi''+\frac12\sin^2\psi\sin2\phi=0.
\end{equation}
The Eq.~\eqref{eq:phi} has a solution
\begin{subequations} \label{eq:onion}
\begin{align}
\label{eq:phi-sol}&\phi^\mathrm{on}(\chi)=\mathrm{am}(x,k),\qquad x=\frac{2\chi}{\pi}\mathrm{K}(k)
\end{align}
where $\mathrm{am}(x,k)$ is Jacobi amplitude\cite{NIST10} and the modulus $k$ is determined by condition
\begin{equation} \label{eq:mu-sol}
2k\mathrm{K}(k)=\pi\sin\psi,
\end{equation}
\end{subequations}
with $\mathrm{K}(k)$ being the complete elliptic integral of the first kind\cite{NIST10}.
The obtained magnetization state is analogous to well known onion-sate with transverse domain walls \cite{Klaui03a}, so we use this name for the solution \eqref{eq:onion}. It should be noted that in the planar limit $\psi\to0$ the onion solution \eqref{eq:onion} is reduced to $\phi^\mathrm{on}=\chi$, that corresponds to uniform magnetization distribution in the Cartesian frame of reference, see Fig.~\ref{fig:onion}b. The solution \eqref{eq:onion} which corresponds to $\psi=\pi/4$ is shown in the Fig.~\ref{fig:onion}c.

\begin{figure}
\includegraphics[width=\columnwidth]{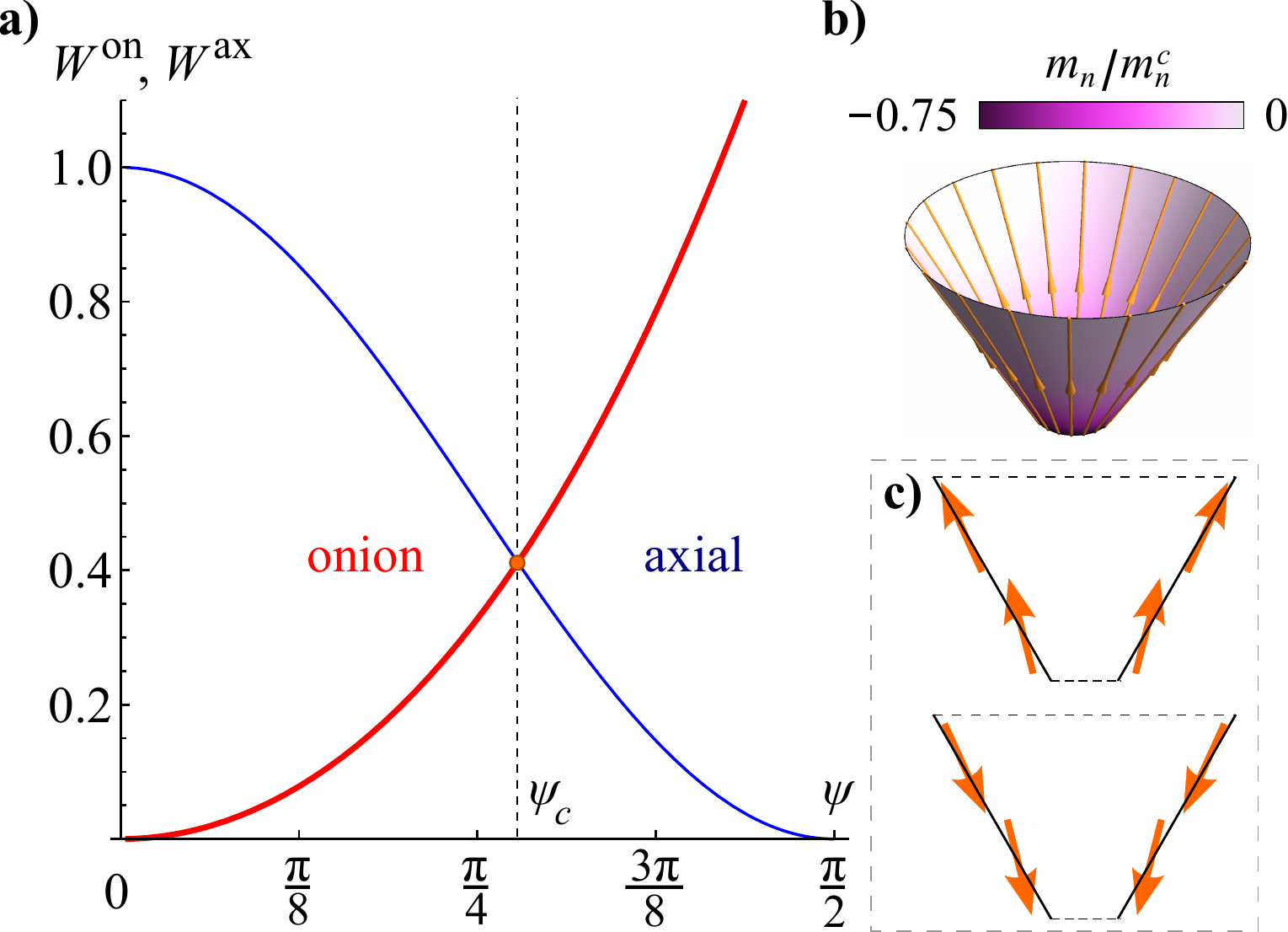}
\caption{(Color online) Energies of the onion (thick line) and axial (thin line) solutions are shown in the plot a). Magnetization distribution of the axial state cone with $\psi=\pi/3$ are shown precisely and schematically on the insets b) and c) respectively. The other notations are the same as in the Fig.~\ref{fig:onion}.}\label{fig:axial}
\end{figure}

To obtain energy of the onion state we substitute the solution \eqref{eq:onion} to \eqref{eq:Eex-phi} and perform the integration over the cone surface in \eqref{eq:Energy-gen}. Finally one can write the onion-state energy as $\mathcal{E}^\mathrm{on}=\mathcal{E}_0(\psi)W^\mathrm{on}$, where
\begin{equation}\label{eq:onion-energy}
W^\mathrm{on}=1-\frac{\sin^2\psi}{k^2}+\frac{4}{\pi}\frac{\sin\psi}{k}\mathrm{E}(k)-2\cos\psi
\end{equation}
with $\mathcal{E}_0(\psi)=2\pi L\ell^2\ln(1+wR^{-1}\cos\psi)/\cos\psi$, and $\mathrm{E}(k)$ being the complete elliptic integral of the second kind\cite{NIST10}. In \eqref{eq:onion-energy} the function $k=k(\psi)$ is implicitly defined by \eqref{eq:mu-sol}. The dependence of energy \eqref{eq:onion-energy} on the generatrix inclination angle $\psi$ is plotted in the Fig.~\ref{fig:axial}a by the thick line. 

On the other hand, the equation \eqref{eq:phi} has another (``axial'') solution $\phi^\mathrm{ax}=\pm\pi/2$ (see Fig.~\ref{fig:axial}b,c) which has the energy $\mathcal{E}^\mathrm{ax}=\mathcal{E}_0(\psi)W^\mathrm{ax}$ with $W^\mathrm{ax}=\cos^2\psi$. Equality of the energies $W^\mathrm{on}(\psi)=W^\mathrm{ax}(\psi)$ determines some critical angle $\psi_c\approx0.8741\approx5\pi/18$ which separates onion ($\psi<\psi_c$) and axial ($\psi>\psi_c$) phases, see Fig.\ref{fig:axial}a.

The obtained evolution of the equilibrium states with the curvature changing (increasing of $\psi$) is an example of a general feature and it can be explained quantitatively as follows. The equilibrium function $\phi$ which minimizes the energy $\mathscr{E}^\mathrm{t}$ appears as a result of competition of three effective interactions: the ``standard'' exchange $\mathscr{E}_{0}^\mathrm{t}=(\nabla\phi)^2$, the effective anisotropy $\mathscr{E}_\mathrm{A}^\mathrm{t}=\vec\Gamma^2$, and effective Dzyaloshinskii-like \cite{Dzyaloshinsky57,*Dzyaloshinsky58} $\mathscr{E}_\mathrm{D}^\mathrm{t}=-2(\nabla\phi\cdot\vec\Omega)$  interactions. For a cone surface one obtains $\mathscr{E}_\mathrm{A}^\mathrm{t}=g^{-1}\sin^2\psi\cos^2\phi$, $\mathscr{E}_\mathrm{D}^\mathrm{t}=-2 g^{-1} \cos\psi \partial_\chi\phi$. The anisotropy term $\mathscr{E}_\mathrm{A}^\mathrm{t}$ dominates for case close to the cylinder ($\psi\to\pi/2$) and it prefers the axial solution $\phi=\pm\pi/2$. On the other hand, the Dzyaloshinskii-like term $\mathscr{E}_\mathrm{D}^\mathrm{t}$ dominates for quasiplanar case $\psi\to0$ and it prefers the solution $\phi\approx\chi$. That agrees with the obtained previously behavior.

It should be noted that appearance of the curvature induced Dzyaloshinskii-like term can explain the observed polarity \cite{Chou07,Curcic08a} and chirality \cite{Vansteenkiste09a} symmetry breaking for magnetic vortices caused by the surface roughness.

Now we estimate the small deviations \eqref{eq:theta-surf} from the obtained tangential solutions originated from the curvature induced effective field $\mathrm{F}^\mathrm{t}$ directed along the normal. For a cone surface  $\mathrm{F}^\mathrm{t}=\sin\psi\sin\phi(2\partial_\chi\phi-\cos\psi)/g$, and therefore one obtains the following values of the normal components $m_n\approx-\vartheta$
\begin{align} \label{eq:normal-comps}
&m_n^\mathrm{on}\approx m_n^c\frac{\sin\psi\,\mathrm{sn}(x,k)}{\left(1+\frac{r}{R}\cos\psi\right)^2}\left[\frac{4}{\pi}\mathrm{K}(k)\mathrm{dn}(x,k) -\cos\psi\right],\nonumber\\
&m_n^\mathrm{ax}\approx\mp m_n^c\frac{\sin\psi\,\cos\psi}{\left(1+\frac{r}{R}\cos\psi\right)^2}
\end{align}
for the onion \eqref{eq:onion} and axial $\phi^\mathrm{ax}=\pm\pi/2$ solutions respectively. Here the normalizing coefficient $m_n^c=\ell^2/(R^2\lambda)$ determines the order of magnitude of the effect, $\mathrm{sn}(x,k)$ and $\mathrm{dn}(x,k)$ are Jacobian elliptic functions \cite{NIST10}. The normal components \eqref{eq:normal-comps} are shown by the color gradient in the Fig.~\ref{fig:onion}c and Fig.~\ref{fig:axial}b for cases of onion and axial solutions respectively. It is interesting to note that $m_n$ decreases with the distance to the cone vertex and it vanishes in the planar limit $\psi\to0$.

The case of strong \underline{easy-normal anisotropy} is much more trivial, the magnetization is oriented along the normal vector (inward or outward the cone surface) up to the small deviations originated from the curvature induced effective magnetic field $\mathrm{F}^\mathrm{n}$ tangential to the surface. The resulting solution \eqref{eq:easy-normal} can be written as $\vartheta=\ell^2\sin2\psi/(|\lambda|g)$ and $\phi=\pm\pi/2$, where the signs ``+'' and ``--'' correspond to inward and outward magnetization orientation respectively. It is interesting to note that for the cylinder surface ($\psi=\pi/2$) the deviation from the normal distribution vanishes.

\begin{figure}
\includegraphics[width=0.8\columnwidth]{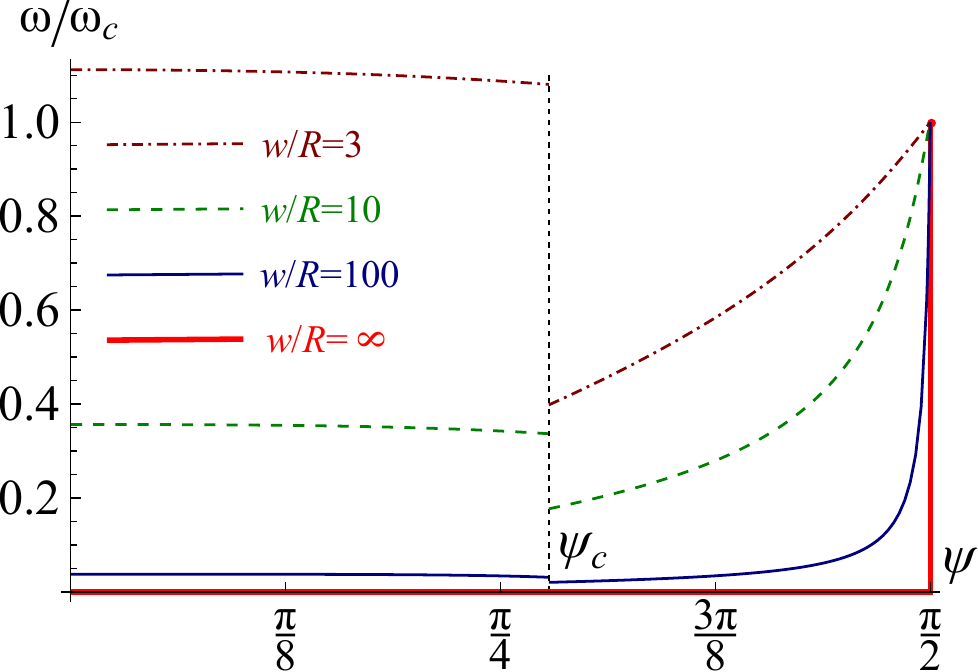}
\caption{The lowest frequencies of linear excitations over the easy-surface ground states depending on the generatrix length $w$ and inclination angle $\psi$.}\label{fig:spectrum}
\end{figure}

As the \textbf{second example} let us consider linear excitations against the obtained equilibrium easy-surface states. Dynamics of a high-anisotropy easy surface shell can be studied using the equation
\begin{equation}\label{eq:phi-dynamics}
\frac{1}{4\lambda}\ddot{\phi}=\ell^2\left[\nabla\cdot(\nabla\phi-\vec\Omega)-\vec\Gamma\cdot\frac{\partial\vec\Gamma}{\partial\phi}\right],
\end{equation}
which follows from the Landau--Lifshitz equation and \eqref{eq:En-easy-surf-all} under the condition $\lambda\gg\ell^2/\mathcal{R}^2$. For the cone surface \eqref{eq:cone-param} the dynamic equation \eqref{eq:phi-dynamics} takes the form
\begin{equation}\label{eq:dynamic}
\frac{g}{4\lambda \ell^2}\ddot\phi = \partial_\chi^2\phi + g \partial_r^2 \phi + \sqrt{g}\cos\psi \partial_r\phi + \frac12 \sin^2\psi\sin2\phi.
\end{equation}
The solution of the Eq.~\eqref{eq:dynamic}, linearized on the background of the onion state \eqref{eq:onion} can be presented in the form
\begin{subequations} \label{eq:Lame-Bessel}
\begin{equation} \label{eq:linear-sep}
\phi(r,\chi,t) \approx \mathrm{am}\left(x,k\right) + e^{i\omega t}\mathrm{P}(\rho)\mathrm{X}(x),
\end{equation}
where $\rho=1+\frac{r}{R}\cos\psi$ and $x$ is defined in \eqref{eq:phi-sol}. By separating variables one can find that the angular part $\mathrm{X}(x)$ satisfies the Lam\'{e} equation \cite{NIST10}
\begin{equation} \label{eq:Lame}
\mathrm{X}'' + \left[h-2k^2\mathrm{sn}^2(x,k) \right]\mathrm{X}=0.
\end{equation}
The periodic solution of \eqref{eq:Lame} which corresponds to the lowest eigenvalue $h=k^2$ \cite{NIST10} coincides (up to the constant) with the following Lam\'{e} function $\mathrm{X}(x) = \mathcal{C}\,\mathrm{Ec}_1^{0}(x,k^2)$. Then the function $\mathrm{P}(\rho)$ appears as the solution  $\mathrm{P}(\rho)= \mathcal{C}_1 \mathrm{J}_0(q\rho) + \mathcal{C}_2 \mathrm{N}_0(q\rho)$ of a zero-order Bessel equation, where $q=\omega/(\omega_c\cos\psi)$ with $\omega_c=2\sqrt{\lambda}\ell/R$. Using the boundary conditions
\begin{equation}\label{eq:bc}
\mathrm{P}'(0)= \mathrm{P}'(\rho_0) = 0
\end{equation}
where $\rho_0=1+\frac{w}{R}\cos\psi$ one can determine the eigenvalues from the following equation $\mathrm{J}_1(q) \mathrm{N}_1(q\rho_0) = \mathrm{J}_1(q\rho_0) \mathrm{N}_1(q)$, whose numerical solution is plotted in the Fig.~\ref{fig:spectrum} for the case $\psi<\psi_c$.
\end{subequations}

Let us analyze now the spin waves on the background of the axial state $\phi^{\mathrm{ax}} = \pm\pi/2$.
Similar to \eqref{eq:Lame-Bessel} one can find that
\begin{equation} \label{eq:linear-sep2}
\phi(r,\chi,t) \approx \pm\frac{\pi}{2} + e^{i\omega t + i\mu\chi}\mathrm{P}(\rho), \qquad \mu\in \mathbb{Z},
\end{equation}
where the radial function $\mathrm{P}(\rho) = \mathcal{C}_1 \mathrm{J}_\nu(q\rho) + \mathcal{C}_2 \mathrm{N}_\nu(q\rho)$, with $\nu=\sqrt{\sin^2\psi + \mu^2}/\cos\psi$. The boundary conditions \eqref{eq:bc} lead to the equation $\mathrm{J}_\nu'(q) \mathrm{N}_\nu'(q\rho_0) = \mathrm{J}_\nu'(q\rho_0) \mathrm{N}_\nu'(q)$ which determines the eigenfrequencies. Its numerical solutions for the lowest mode $\mu=0$ are plotted in the Fig.~\ref{fig:spectrum} for the case $\psi>\psi_c$. As well as in the previous case, the lowest frequency becomes arbitrary small with the cone size increasing. Nevertheless it is not so for the cylinder surface where the lowest frequency is fixed and it is equal to $\omega_c$. The case of cylinder ($\psi=\pi/2$) should be considered separately starting from the Eq.~\eqref{eq:dynamic}, whose linear solution against the axial state has the form $\phi=\pm\pi/2+\mathcal{C}e^{i(\omega t+\mu\chi+\kappa r)}$ with $\kappa$ being the wave vector along cylinder axis. The corresponding dispersion relation reads $\omega=\omega_c\sqrt{1+\mu^2+R^2\kappa^2}$. Existence of a gap in spectrum of the cylindrical magnetic shell was already predicted theoretically \cite{Gonzalez10} and checked by numerical simulations \cite{Yan11a}.

In summary we dare to make some remarks about possible perspectives of development of the curvilinear magnetism area. On the one hand, new effects which occur due to the curvature are expected to be of order of magnitude $\ell/\mathcal{R}$. Since the typical values are $\ell\lesssim10$ nm and $\mathcal{R}>10^2$ nm the curvature effects are expected to be small. Nevertheless we can formulate several perspective directions in the studying of the curvilinear nanomagnets: (i) Topological effects. Equilibrium states of a curvilinear shell with high easy-surface (or easy normal) anisotropy are determined by topological properties of the surface, e.g. vortices on a spherical shell appears as a result of the hairy ball theorem \cite{Eisenberg79}. (ii) Nonlocal effects. Being of small magnitude some curvilinear effects can be spatially nonlocal, e.g. the deformation of the in-surface structure of a vortex on spherical shell \cite{Kravchuk12a}. Such nonlocal magnetization deformations can modify the interaction between nonlocalized magnetization structures (e.g. vortices or antivortices).  (iii) Chiral effects. The curvature can remove the chirality degeneration, which is typical for planar systems, e.g. chirality-polarity coupling on a spherical shell \cite{Kravchuk12a}, or breaking of chirality symmetry in vortex domain wall on a cylindrical tube \cite{Otalora12,Yan12,Otalora13}.

In this regard, we believe that the proposed general expression for the exchange energy \eqref{eq:main-formula} for an arbitrary curvilinear surface opens a new direction of the theoretical study of the curvilinear magnetic nanoshells.

The authors thank D. Makarov for stimulating discussions and acknowledge the IFW Dresden, where part of this work was performed, for kind hospitality. This work was partially supported by DFG project MA 5144/3-1, and by the Grant of President of Ukraine for support of researches of young scientists (Project No~GP/F49/083).

%
%
%
%

\end{document}